# Frequency-Resolved Forward Capacitance in GaN-based LEDs under different currents

Yuchen Li[1], Zhizhong Chen[1,2,3,*], Chuhan Deng[1], Boyan Dong[1], Daqi Wang[1], Zuojian Pan[1], Haodong Zhang[1], Jingxin Nie[1], Weihua Chen[1], Fei Jiao[1,4], Xiangning Kang[1], Qi Wang[2], Guoyi Zhang[1,2], Bo Shen[1,3], Wenji Liang[5]

1 State Key Laboratory for Artificial Microstructure and Mesoscopic Physics, School of Physics, Peking University, Beijing 100871, China

2 Dongguan Institute of Optoelectronics, Peking University, Dongguan, Guangdong 523808, China

3 Yangtze Delta Institute of Optoelectronics, Peking University, Nantong, Jiangsu 226000, China

4 State Key Laboratory of Nuclear Physics and Technology, School of Physics, Peking University, Beijing 100871, China

5 AET Displays Limited Co., Ltd, Dongguan, Guangdong, China

*E-mail: zzchen@pku.edu.cn

Abstract: This study establishes a unified framework for interpreting dynamic capacitive responses in InGaN-based light-emitting diodes (LEDs) through forward-bias capacitance-voltage-frequency spectroscopy. A hybrid impedance model integrating series RL components and parallel C-G networks was developed to resolve distinct frequency-dependent capacitive regimes. The low-frequency regime (<1 kHz) is governed by interfacial capacitance with characteristic reciprocal frequency dependence, while the mid-frequency range (10–300 kHz) demonstrates carrier diffusion and recombination dynamics. At MHz frequencies, negative capacitance manifests due to delayed carrier emission mediated by deep-level traps. The model achieved sub-1% fitting errors ($R^2 > 0.99$) across a broad bandwidth (10 kHz–6.4 MHz), conclusively attributing negative capacitance to intrinsic trap processes rather than extrinsic artifacts. Critical advances include quantum well cap thickness modulation reducing mid-frequency capacitance by ~30% and the dominance of trap-mediated inductance over parasitic contributions by three orders of magnitude. This framework resolves persistent controversies in LED impedance interpretation. By bridging semiconductor physics with device engineering, this methodology provides essential tools for designing next-generation optoelectronic systems requiring ultralow-latency operation and precise charge-state control.

## Introduction

InGaN-based light-emitting diodes (LEDs) have emerged as cornerstone technologies for next-generation displays, where their inherent nanosecond-scale response theoretically enables unprecedented refresh rates and motion clarity [1,2]. Yet current architectures fail to fully exploit the photoelectric conversion potential of quantum well active regions, with device-level response times often limited by capacitive effects—particularly the enigmatic negative capacitance (NC) under forward bias [3]. The frequency-dependent characteristics of negative capacitance (NC) phenomena in semiconductor devices, documented over decades [4-11], continue to pose significant interpretative challenges. Current research exhibits fundamental discrepancies in mechanistic understanding: Shim et al. [12] ascribe NC artifacts to series resistance effects through capacitance-voltage (C-V) characterization results at 1 MHz, while

Feng et al. [13-16] attribute the phenomenon to carrier lifetime variations in active regions. Although these models demonstrate self-consistency within their respective experimental frameworks, their dependence on limited frequency sampling (≤3 discrete points) fails to capture essential dispersion characteristics. Notably, while Ershov et al. derived the negative capacitance effect in semiconductor devices by invoking an exponentially decaying transient current [17], a reasonable explanation for the physical origin of this current remains absent in their framework. This gap in understanding undermines the mechanistic clarity of the negative capacitance phenomenon proposed in the study, highlighting a critical unresolved question in the underlying physics.

To address these unresolved controversies, we present a comprehensive experimental framework integrating broadband frequency-swept capacitance spectroscopy (40 Hz–6.4 MHz) with a hybrid circuit modeling approach. Our modeling framework integrates two complementary elements: series RL components that quantitatively capture trap-state carrier emission dynamics, coupled with parallel C-G network configurations that explicitly describe active-region behavior. This synergistic approach enables identification of three operationally distinct capacitive regimes, characterized respectively by metal-semiconductor interface dominance below 1 kHz, active-region diffusion-recombination mediation near 100 kHz, and trap-mediated carrier emission predominance above 3 MHz. Crucially, the combined experimental and theoretical analysis establishes that negative capacitance manifestations originate from intrinsic trap-state emission processes rather than extrinsic measurement artifacts. Our findings provide a unified framework explaining frequency-bias interdependence of LED capacitance, enabling predictive design rules. This methodology offers new insights into response characteristics of optoelectronic devices.

## Experimental methods

GaN-based LED structures were epitaxially grown on 2-inch sapphire substrates using an AIXTRON CRIUS MOCVD system equipped with closed-loop gas flow control and multi-zone thermal regulation. The substrates were initially cleaned at 1050°C under hydrogen atmosphere, followed by sequential growth of a GaN buffer layer, undoped GaN, and n-type GaN. The active region comprised an 8-period superlattice structure and a single quantum well. The capping layer (~0.75 nm) of LED A was deposited at 830°C with triethylgallium (TEGa) flow rate of 90 mL/min for 90 s. Notably, trimethylaluminum (TMAl) at 50 mL/min was introduced during the final capping stage for LED B samples (AlGaN capping ~1 nm). The epitaxial structure was completed by growing low-temperature p-GaN, conventional p-GaN, and a p-InGaN contact layer.

Device fabrication involved UV photolithography to pattern isolated light-emitting units, followed by inductively coupled plasma etching using $Cl_2$/$BCl_3$ to expose the n-GaN layer. Electrodes (p-Ni/Au, n-Ti/Al/Ni/Au) were deposited via thermal evaporation, with subsequent rapid thermal annealing under nitrogen ambient to optimize doping activation and contact resistance.

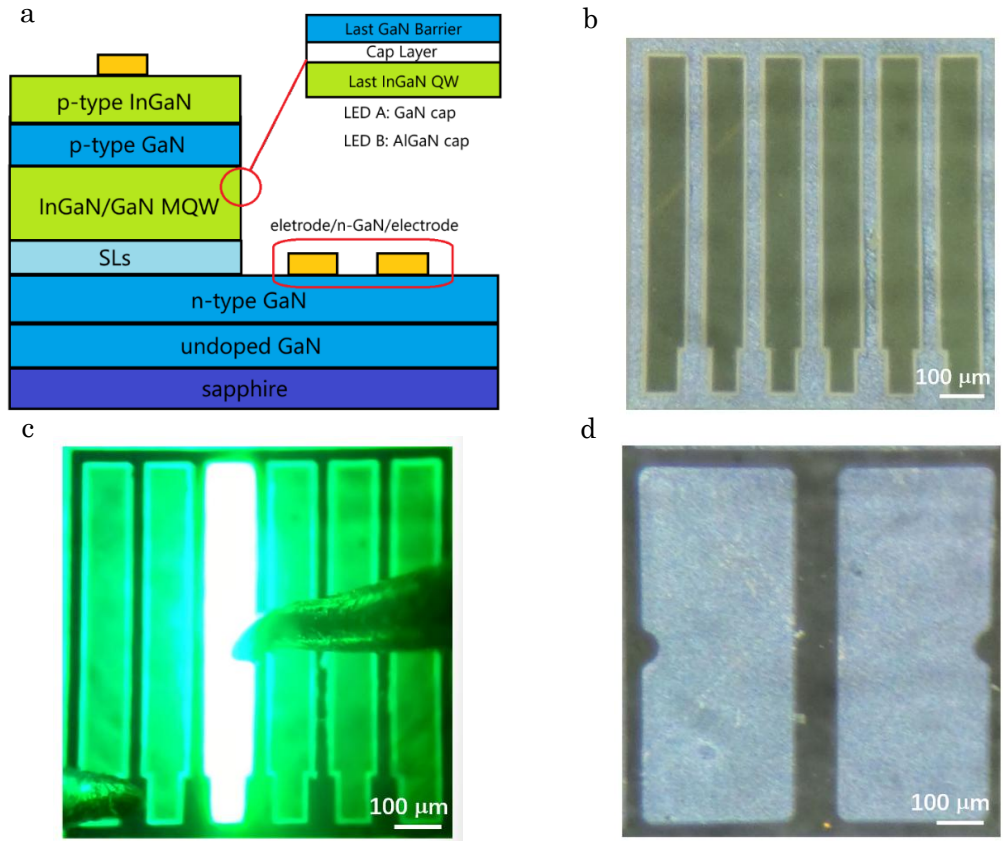

Figure 1. LED sample schematic and optical microscopy images: Schematic illustration of the device structure (a); Optical microscopy image of the LED chip (b)&(c); Optical microscopy image of the simplified electrode-nGaN-electrode structure at the alignment mark,

Capacitance characterization was performed using an Agilent 4294A impedance analyzer (40 Hz–110 MHz, ±0.08% accuracy) in parallel $C_p$- $g$ mode. The system was calibrated with open/short/load compensation to eliminate parasitic effects. Measurements employed a 25 mV AC excitation under DC bias, maintaining small-signal conditions ($V_{ac}$< kT/q), with frequency sweeps divided into three ranges: low frequency (40 Hz–1.04 kHz): 101 points to mitigate drift during prolonged integration, mid-frequency (1–801 kHz): 801 points, high frequency (800 kHz–6.4 MHz): 801 points

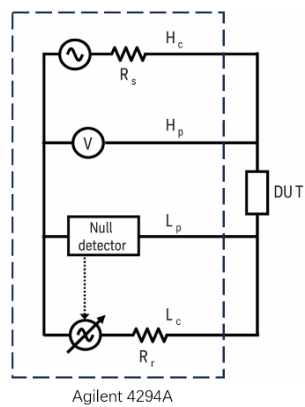

Figure 2. Schematic diagram of capacitance testing circuit

The parallel capacitance-conductance model decomposed device impedance into orthogonal components: $Im(I_{ac}) = \omega C_p V_{ac}$, $Re(I_{ac}) = gV_{ac}$, where $\omega$ denotes angular frequency. Bandwidth optimization (Level 5: 200 ms/point) enhanced high-frequency resolution critical for analyzing interface states and charge storage behavior. All measurements were conducted at 293 K.

## Results and Discussion

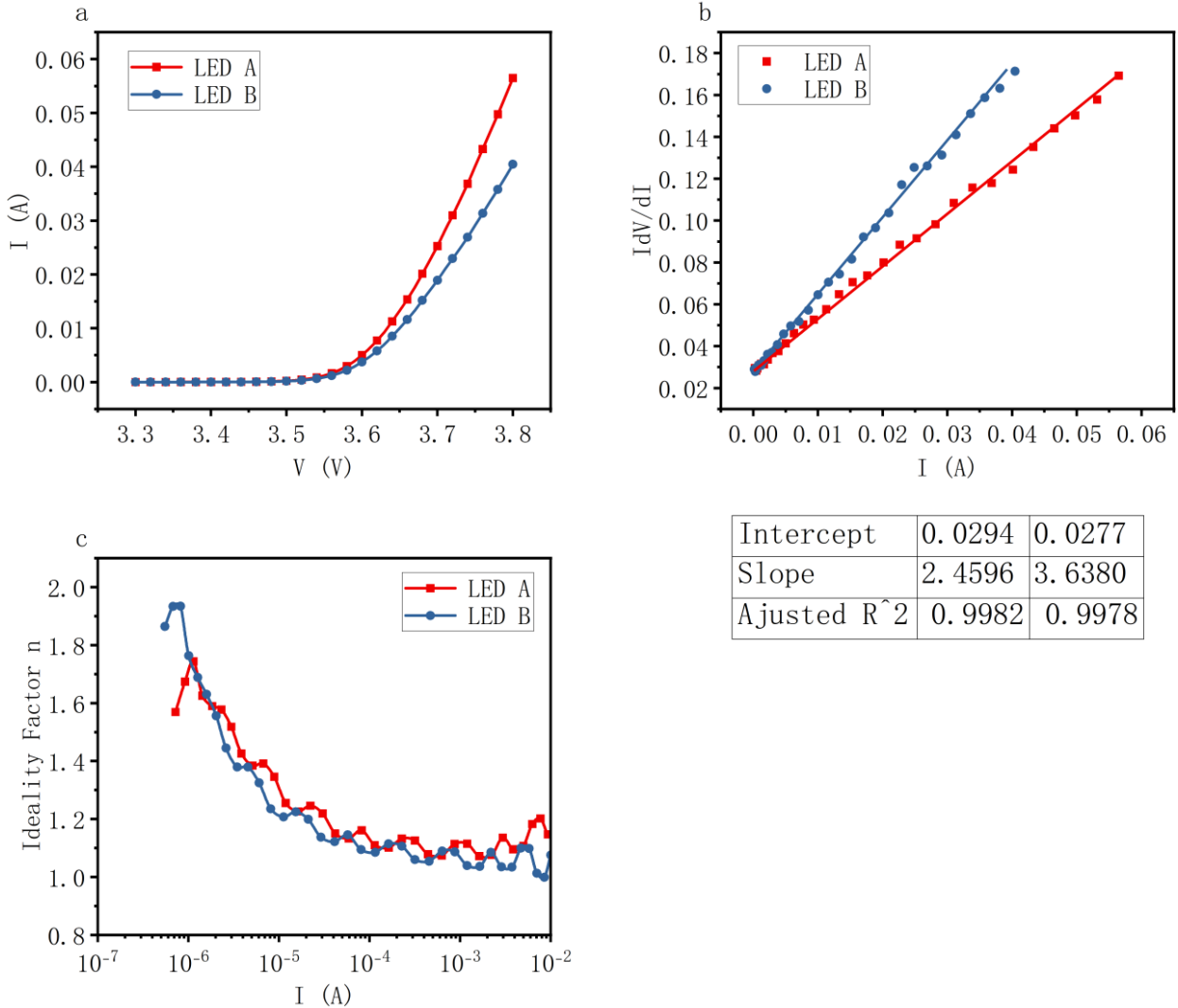

Figure.3 current-voltage (I-V) characteristics (a), extracted series resistance (b) and ideality factor of the LEDs (c).

Current-voltage (I-V) characterization reveals distinct transport properties between devices. The extracted series resistances, obtained through linear regression of the IdV/dI vs. I plot, measure 2.46 Ω for LED A and 3.64 Ω for LED B. This increase in series resistance for LED B likely originates from thicker capping layer. The ideality factor (n) exhibits strong current dependence (Fig. 3c): when I≈1 μA, n≈1.7, indicating dominant non-radiative recombination, while when I≈1 mA, n approaches 1.1, signifying transition to radiative recombination current domination.

The differential capacitance in the sub-kHz regime demonstrates characteristic reciprocal frequency dependence ($C \propto 1/f$), as evidenced in Figure 4. Quantitative analysis reveals a systematic increase in absolute capacitance magnitude with decreasing frequency, culminating in pronounced negative capacitance values below 1 kHz. This anomalous behavior fundamentally diverges from established semiconductor capacitance mechanisms (e.g., depletion-layer modulation or diffusion capacitance), strongly indicating unconventional charge dynamics at play.

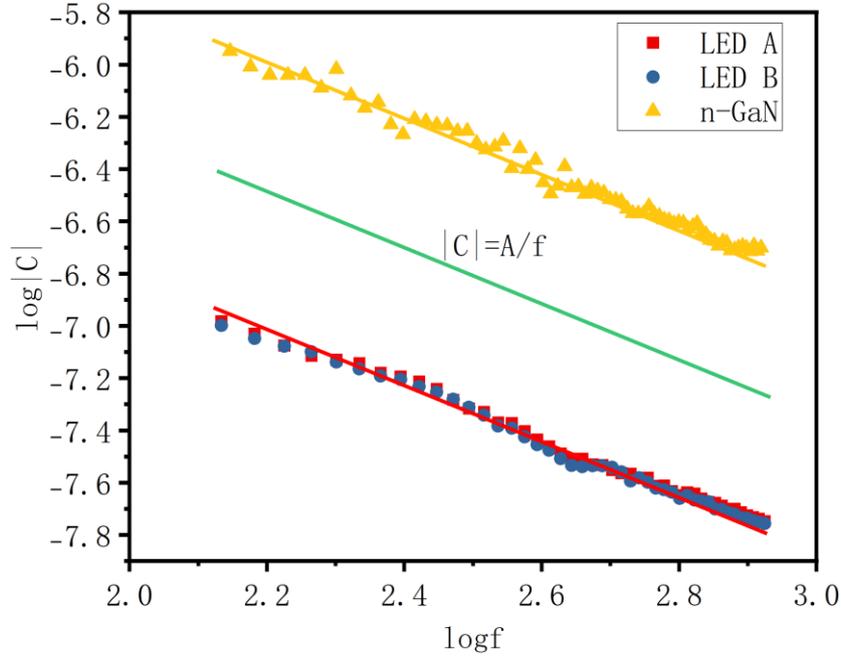

Figure.4 Low-frequency C-f plots in double-logarithmic coordinates of LED A, LED B and simplified n-GaN structure (current density@40 A/cm²)

To mechanistically decouple interface contributions from bulk phenomena, we engineered a prototypical test structure (metal/n-GaN/metal) fabricated using optical lithographic alignment markers (Figure 1, red dashed region). This simplified architecture intentionally eliminates quantum confinement effects, p-n junction interfaces, and minority carrier recombination pathways. Strikingly, the simplified structure's low-frequency capacitance dispersion (Figure 4) demonstrates fundamental consistency with full LED devices through three critical aspects: maintained inverse frequency scaling relationship, identical negative capacitance onset below 1 kHz, and capacitance magnitudes of equivalent order. These collective observations provide conclusive evidence that the governing mechanism resides in metal-semiconductor interfacial phenomena rather than bulk charge transport or junction-related processes.

The observed $1/f$ scaling law presents a fundamental contradiction to classical Debye relaxation theory predicated on discrete trap states. We resolve this paradox through a continuum approximation framework accounting for distributed trap lifetimes. Whereas individual traps with characteristic lifetime τ produce:

$$C(\omega) = a \int_0^\infty \exp\left(-\frac{t}{\tau}\right) \cos\omega t \, dt = \frac{a\tau}{1+\omega^2\tau^2} \quad (1)$$

a logarithmic lifetime distribution ($g(\tau) \propto 1/\tau$) yields:

$$C(\omega) \propto \int_0^\infty \frac{\tau}{1+\omega^2\tau^2}\frac{1}{\tau}\,d\tau = \frac{\pi}{2\omega} \qquad (2)$$

This The model matches experimental data, suggesting long-lifetime traps ($\tau \sim 1$ ms) influence low-frequency capacitance. GaN-based devices, like Pt/AlGaN/GaN diodes [18], show hydrogen-induced capacitance increases below 1 kHz, where interfacial effects dominate. The authors suggested this could relate to H-induced dipolar reconfiguration: sub-1 kHz AC fields might switch dipole polarity at metal-semiconductor interfaces, enhancing dielectric response. This frequency-dependent modulation aligns with modeled trap dynamics ($\tau \sim 1$ ms), implying a possible link between "long-lifetime traps" in 1/f dispersion and metastable H-dipole states needing millisecond relaxation.

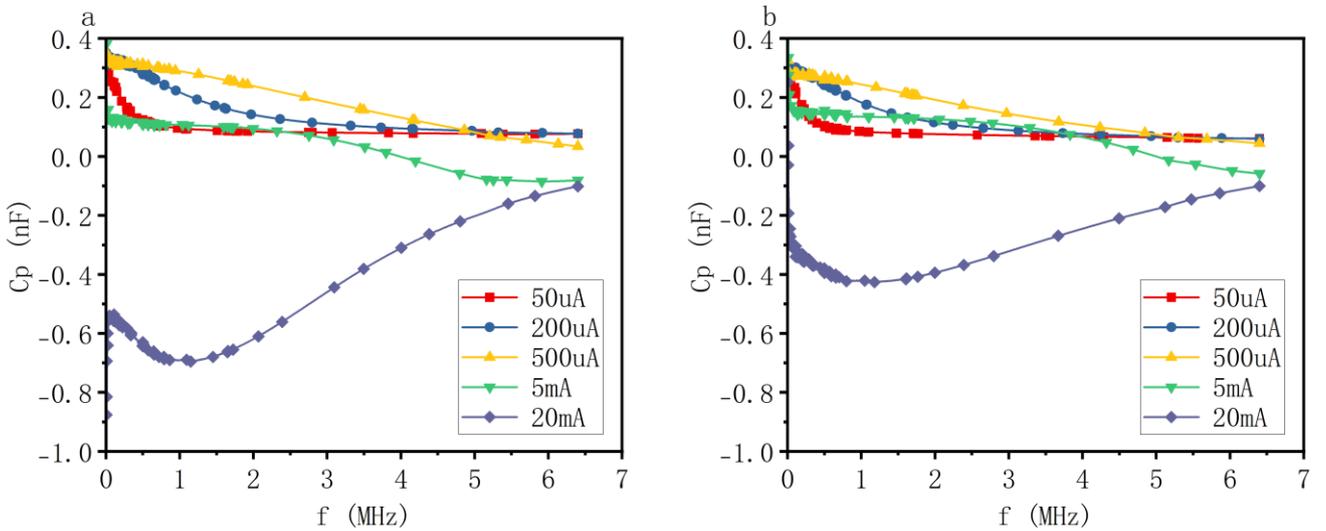

Figure.5 Mid-to-high frequency C-f characteristics of LED A and LED B

While interfacial dynamics dominate the sub-1 kHz regime, the extended frequency range (1 kHz–6.4 MHz) reveals intricate capacitive-inductive coupling mechanisms. As demonstrated in Figure 5, GaN-based LED devices exhibit frequency-dependent capacitance behavior modulated by carrier injection levels, manifesting three characteristic phases. The 1 kHz–20 kHz range maintains residual interfacial capacitance influence, particularly under near-zero injection conditions where classical p-n junction behavior prevails. Systematic analysis of injection-dependent responses uncovers nonlinear evolution: from minimal injection levels, capacitance magnitude increases monotonically with current density, reaching maximum values near 0.5 mA in the 1 MHz regime. Beyond this critical threshold, two concurrent phenomena manifest—a gradual transition from concave to convex curvature at 1 MHz accompanied by emergent negative capacitance characteristics at the upper frequency limit (6.4 MHz). These frequency-selective responses exhibit characteristics analogous to LC series resonance phenomena, when capacitive dominance at lower frequencies inverts to inductive behavior at elevated frequencies. This compelling correspondence suggests the existence of intrinsic equivalent inductive elements within the device architecture, potentially arising from delayed carrier emission processes.

To address this, the AC small-signal response of GaN-based LEDs is analyzed using a series model, focusing on the imaginary component of differential impedance. At frequencies >1 kHz, particularly under low current, the devices exhibit LC-like series resonance. The series model proves more intuitive than the parallel capacitance-conductance ($C_p$- $g$) model commonly used in prior studies [3-17].

$$\tilde{Z} = \frac{1}{\tilde{Y}} = \frac{1}{g + i\omega C_p} = \frac{g - i\omega C_p}{g^2 + \omega^2 C_p^2} \qquad (3)$$

Where $g$ is conductance, $C_p$ is parallel capacitance, and $\omega$ is angular frequency. Separating real and imaginary components yields:

$$Re(\tilde{Z}) = \frac{g}{g^2 + \omega^2 C_p^2} = R_s; \ Im(\tilde{Z}) = \frac{-\omega C_p}{g^2 + \omega^2 C_p^2} = \omega L_s \qquad (4)$$

Here, a positive capacitance corresponds to a negative inductance, reflecting the capacitive-inductive duality.

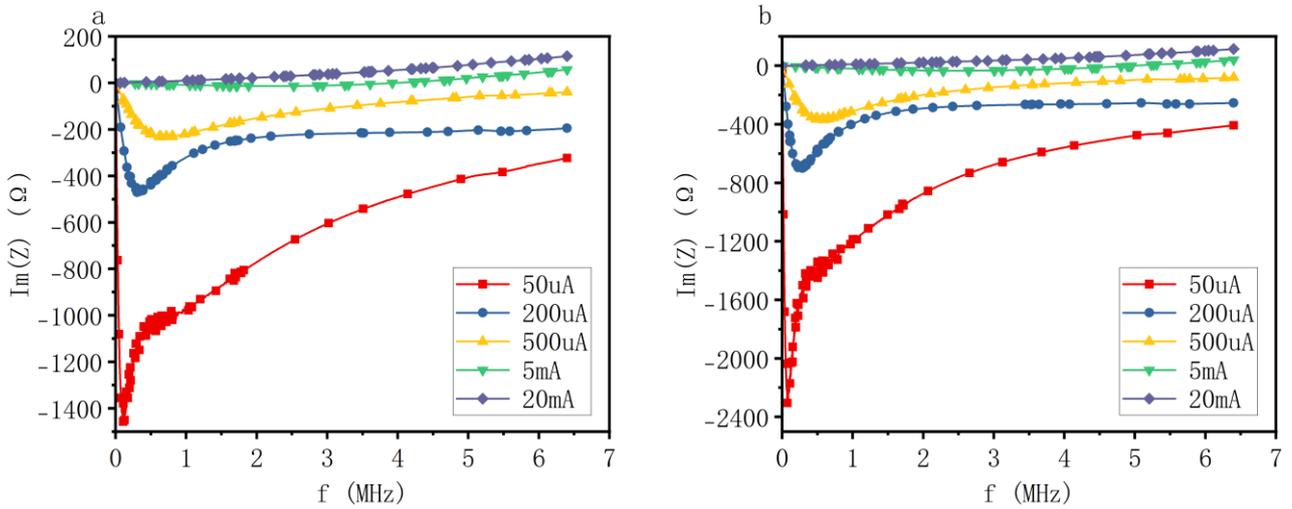

Figure.6 Experimental profile of frequency-dependent imaginary impedance (Im(Z)) under forward injection

Figure 6 demonstrates distinctive frequency-modulated imaginary impedance evolution under forward bias operation, revealing three principal characteristics through systematic analysis. The high-frequency regime exhibits progressively increasing Im(Z) magnitudes that directly correlate with observed negative capacitance phenomena, indicative of inductive coupling mechanisms. A critical transition emerges near 1 mA injection current where Im(Z) undergoes polarity inversion from negative to positive values, exhibiting spectral signatures analogous to LC resonance characteristics. Concurrently, the sub-MHz domain displays nonlinear dispersion marked by an initial linear descent to a characteristic minimum followed by gradual recovery, forming a distinctive low-frequency dip profile.

These multifrequency responses are mechanistically attributed to the interplay between trap-mediated carrier emission kinetics, active-region diffusion dynamics, and recombination process modulation. Our proposed unified equivalent circuit model (Fig. 7) successfully reconstructs the complete impedance spectrum through three functionally integrated components: a contact interface subsystem comprising parallel capacitance ($C_c$) and conductance ($g_c$) elements that characterize metal-semiconductor interfacial properties, an active-region network incorporating junction capacitance ($C_a$) with conductance ($g_a$) parameters that collectively describe diffusion and recombination processes, and a high-frequency trap module featuring series-connected inductance ($L_{trap}$) and resistance (R_s) elements that physically capture delayed carrier emission dynamics. This tripartite architecture comprehensively addresses both interfacial charge storage mechanisms and bulk transport phenomena while quantitatively reproducing the characteristic negative capacitance signatures observed experimentally.

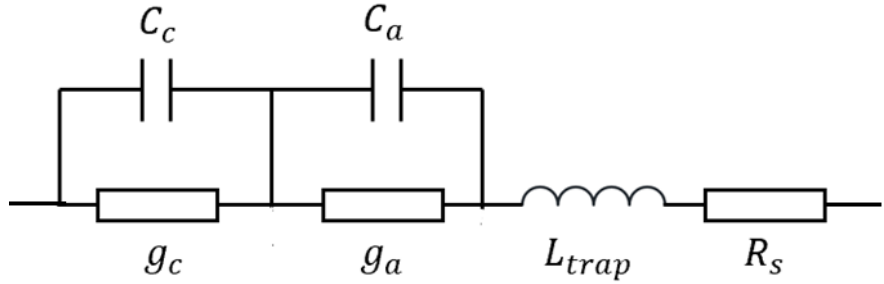

Figure.7 Equivalent circuit for AC small-signal frequency response of GaN-based LEDs

The imaginary impedance is expressed as:

$$Im(Z) = Im(Z_c) + Im(Z_a) + \omega L_{trap} \tag{5}$$

Where $Zc = 1/(g_c + i\omega C_c)$. Contact impedance contributions (Im($Zc$)~0.1 Ω) are negligible compared to the total impedance (~100 Ω).

Using Equation:

$$Im(Z) = -\frac{\omega C_a}{g_a^2 + \omega^2 C_a^2} + \omega L_{trap} \tag{6}$$

the experimental Im(Z)-$f$ data (Figure 6) are fitted to extract $C_a$, $g_a$, and $L_{trap}$ (Figure 8&9). Key findings include: $L_{trap}$ increases with injection current, saturating at ~5μH for $I$>10 mA; $C_a$ and $g_a$ scale with current, aligning with the diffusion-recombination model.

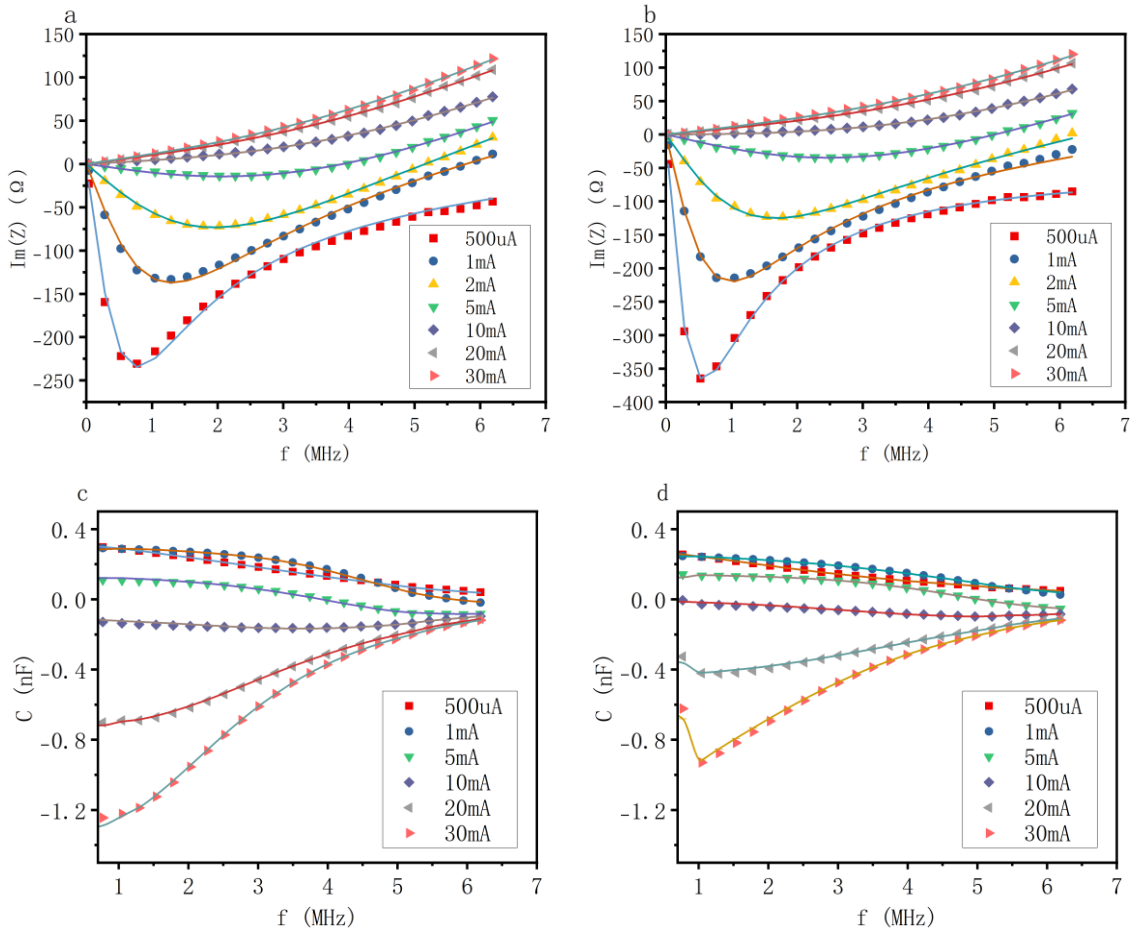

Figure.8 Fitting results of the equivalent circuit model for Im(Z) vs experimental data (LED A-a, LED B-b) and capacitance (LED A-c, LED B-d) in the equivalent circuit

LED A (GaN cap layer) and LED B (AlGaN cap layer) exhibit distinct $C_a$ and $g_a$ behaviors (Figure 9): Lower $g_a$ and $C_a$ in LED B. This could be attributed to its stronger electron blocking capability, which reduces carrier overflow and non-radiative recombination, while possibly more effectively passivating surface traps and positively influencing the internal stress distribution. These factors collectively improve the carrier recombination environment in the active region and enhance overall device performance.

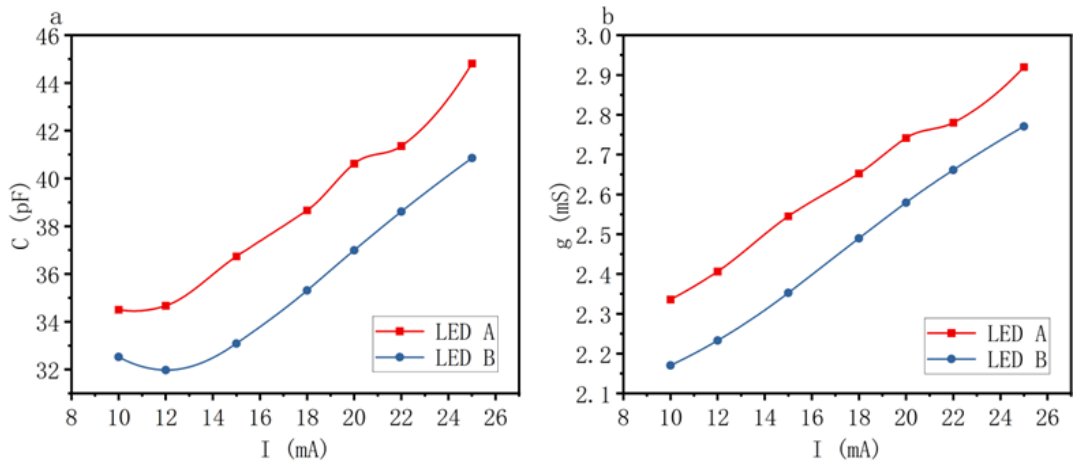

Figure.9 Fitting results of $C_a$ (a) and $g_a$ (b) in the equivalent circuit

While the extracted equivalent inductance (L≈5 $\mu$H) exceeds parasitic contributions (L ~nH), conventional packaging-induced inductance cannot account for its strong response effect. This orders-of-magnitude discrepancy necessitates a carrier-mediated origin. The existing studies on the effect of trap states on the electrical characteristics of devices mainly involve the change of verse-bias capacitance or capacitance hysteresis [19-25]. Xu et al. proposed a trap state carrier release model where stored charges generate effective inductance by introducing an additional "collisional loss" term proportional to injection current density (*J*) into the Shockley-Read model [26-27]:

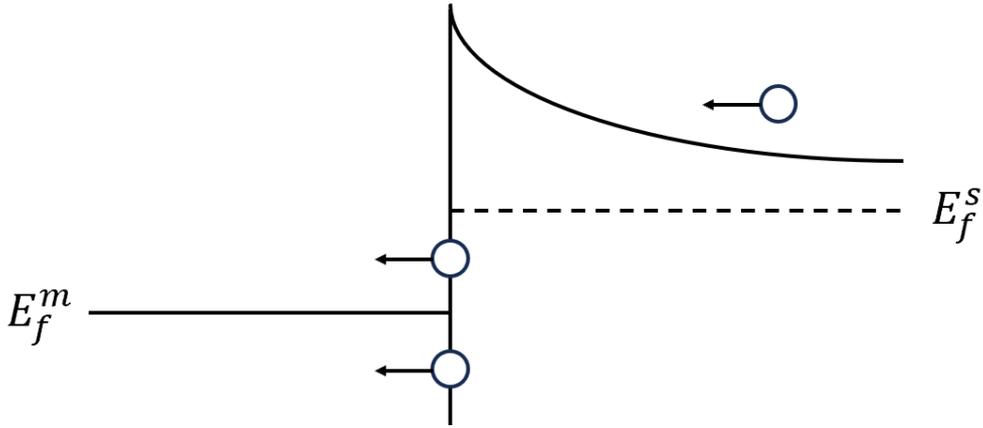

Figure.10 Schematic diagram of the carrier emission mechanism from interface states under injection current

At interfaces (Figure. 10), electrons crossing the barrier under forward bias possess sufficient kinetic energy to collide with trapped electrons, ejecting them from trap/localized states.

The occupation probability *F(t)* evolves as [26-27]:

$$\frac{dF(t)}{dt} = c_n[1 - F(t)]n_s(t) - e_nF(t) - c_pF(t)p_s(t)$$
$$+ e_p[1 - F(t)] - \frac{F(t) - F_m}{\tau_m} - \frac{a_n}{4}F(t)n_s(t) \quad (7)$$

Where $c_n$ ($c_p$) and $e_n$ ($e_p$) represent the electron (hole) capture coefficients and electron (hole) emission constants, respectively; $n_s(t)$ and $p_s(t)$ are the concentrations of free electrons and holes on the semiconductor surface at time $t$; $F_m$ and $\tau_m$ denote the Fermi function on the metal side and the relaxation time, respectively. Under forward bias, electrons crossing the barrier fill the empty interface states. However, due to their additional kinetic energy, when colliding with the trapped electrons in the states, if the binding energy is lower than the barrier energy, the trapped electrons can be ejected.

Under AC modulation, the occupation response is derived as [23-24]:

$$\delta f_1 = \frac{4\sigma_n \delta j/q}{\tau_1\left[(4\sigma_n + \sigma_i)J/q + \frac{1}{\tau_1}\right]\left[(4\sigma_n + \sigma_i)J/q + \frac{1}{\tau_1} + i\omega\right]} = \frac{A_1}{1 + i\omega\tau_{eff1}}$$

$$\delta f_2 = -\frac{\sigma_i \delta j/q}{\tau_2\left[(4\sigma_n + \sigma_i)J/q + \frac{1}{\tau_2}\right]\left[(4\sigma_n + \sigma_i)J/q + \frac{1}{\tau_2} + i\omega\right]} = -\frac{A_2}{1 + i\omega\tau_{eff2}} \quad (8)$$

Here, $\tau_1$ and $\tau_2$ represent lifetimes for states above/below the Fermi level, $\tau_{eff} = \tau/[1 + (4\sigma_n + \sigma_i)J\tau/q]$. Converting to series $R_s - L$ model parameters:

$$L_2 = \frac{kTC_i}{q\sigma_i} \frac{\Delta V}{\delta j} \frac{1 + (4\sigma_n + \sigma_i)J\tau_2/q}{J} \quad (9)$$

At high injection ( $(4\sigma_n + \sigma_i)J\tau/q \gg 1$ ), $L$ asymptotically approaches a constant, consistent with experimental trends in Figure.6 and Figure.8. States below the Fermi level contribute positive $R_s$ and $L$ (or negative capacitance), while those above yield negative values. In GaN-based LEDs, high-frequency responses are dominated by trap/localized states below the Fermi level, with negative capacitance arising from carrier detrapping and redistribution. The derived model explains the current-dependent inductance observed in Figure.6.

The variable temperature test results (Figure. 11) further elucidate the dynamic behavior of these states. As temperature decreases, the observed reduction in inductance L indicates strong temperature dependence in the carrier emission process from the states. At lower temperatures, the reduced probability of carrier emission from states alters carrier distribution and dynamic response within the device, thereby modifying its impedance characteristics. This temperature dependence provides critical insights into both the energy distribution of trap/localized states and their capture/emission kinetics, while further corroborating the pivotal role of these states in governing the electrical properties of GaN-based LEDs.

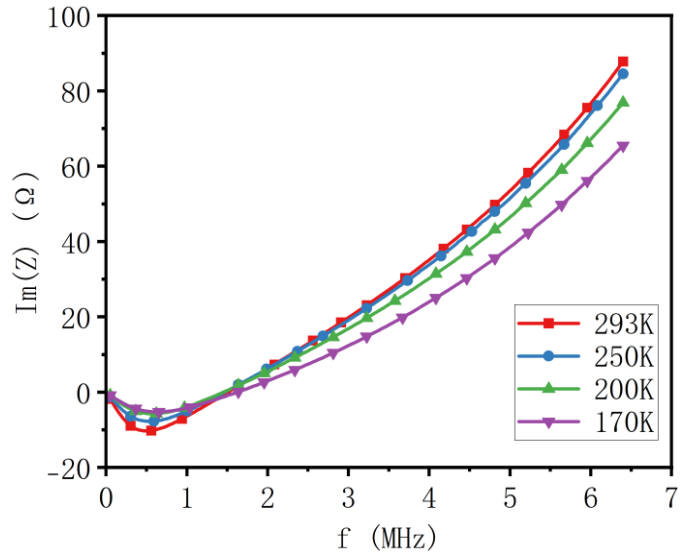

Figure 11. Im(Z)-frequency relationship diagram of LED under constant injection current (1 mA) at different temperatures, demonstrating that the equivalent inductance L decreases with decreasing temperature.

# Conclusion

This investigation establishes a novel analytical framework for interpreting the dynamic capacitive response of GaN-based LEDs through forward-bias capacitance-voltage-frequency spectroscopy. The methodology enables clear resolution of three distinct frequency-dependent operational regimes: sub-kHz dynamics governed by interfacial phenomena exhibiting characteristic reciprocal frequency dependence ($C \propto 1/f$), mid-frequency behavior (10–300 kHz) dominated by recombination processes, and MHz-range negative capacitance manifestations originating from carrier emission processes mediated by states below the Fermi level. This system provides critical insights into the fundamental interplay between interfacial charge relaxation, bulk transport kinetics, and state-mediated transient responses in wide-bandgap optoelectronic devices. The proposed hybrid series-parallel impedance model transcends conventional analysis by unifying these phenomena under a single formalism, achieving sub-1% fitting errors ($R^2 > 0.99$) across five decades of frequency (40 Hz–6.4 MHz).

Two transformative advances emerge: First, the identification of quantum well cap thickness as a capacitance-tuning knob—reduces active-region conductance and capacitance by ~10%, suppressing mid-frequency dispersion. Second, the revelation of state emission mediated negative capacitance that dominates over parasitic contributions by three orders of magnitude, redefining high-frequency device operation principles.

While our current experimental configuration is frequency-limited to 6.4 MHz by conventional probe, the developed methodology successfully resolves long-standing controversies in LED impedance interpretation. Future research will pursue three strategic research thrusts: advancing temperature-dependent impedance mapping across cryogenic to 400 K regimes to disentangle thermal activation processes from intrinsic carrier dynamics, developing advanced probing techniques for GHz-range frequency characterization to access ultrafast kinetics, and extending cross-material platform validation studies to β-$Ga_2O_3$ and AlN-based heterostructures. This analytical framework establishes a critical nexus between fundamental semiconductor physics and practical device engineering, providing an indispensable toolkit for designing next-generation optoelectronic systems requiring ultralow-latency operation and precision charge-state control.

# Acknowledgment

This work was supported by National Key Research and Development Program (2023YFB4604400); National Natural Science Foundation of China (62174004, 61927806); and Guangdong Basic and Applied Basic Research Foundation (2020B1515120020).